\documentclass[11pt, a4paper]{article}

\usepackage{amsmath}
\usepackage{amssymb}
\usepackage{graphicx}
\usepackage{hyperref}
\usepackage{geometry}
\usepackage[T1]{fontenc}

\title{Microscopic Structure of Random 3-SAT: \\A Discrete Geometric Approach to Phase Transitions and Algorithmic Complexity}
\author{Yongjian Zhan}
\date{\today}

\begin{document}

\maketitle

\begin{abstract}
The structural phase transitions and computational complexity of random 3-SAT instances are traditionally described using thermodynamic analogies from statistical physics, such as Replica Symmetry Breaking and energy landscapes. While providing profound macroscopic insights, these theories lack a discrete microscopic structure. In this paper, we propose a complementary, strictly discrete geometric model that maps these phenomena directly to the combinatorial topology of an $N$-dimensional Boolean hypercube. By defining the problem space purely through valid solutions rather than abstract energy states, we establish deterministic mechanics for clustering and freezing, driven by the progressive elimination of vertices and Hamming distance bridges. Furthermore, we derive absolute structural boundaries for 3-SAT, identifying a minimal unsatisfiability limit at constraint density $\alpha = \frac{8}{N}$ populated by at least $\frac{N(N-1)(N-2)}{6}$ distinct unsatisfiable cores, and a maximal satisfiability limit at $\alpha = \frac{7}{6}(N-1)(N-2)$ populated by $2^N$ maximal satisfiable instances. These combinatorial extremes mathematically elucidate why the average-case Satisfiability Threshold Conjecture holds only ``almost surely.'' Finally, we apply this topological framework to explain the ``easy-hard-easy'' algorithmic complexity curve. We demonstrate that the efficiency of Depth-First Search is governed by the geometric transition from an abundance of valid search paths (the under-constrained easy phase) to a high density of structurally ``removed variables'' that force immediate contradictions (the over-constrained easy phase). This microscopic perspective bridges theoretical phase transitions with the concrete mechanics of complete search algorithms.
\end{abstract}

\vspace{1em}

\newpage

\section{Introduction}

The study of random 3-SAT instances has long served as a benchmark for understanding computational complexity and algorithmic performance \cite{Biere2021, Mezard2009, Monasson1999}. As the constraint density---defined as the ratio of clauses to variables, $\alpha=\frac{M}{N}$---increases, the geometry of the solution space undergoes dramatic structural phase transitions. Historically, the theoretical framework explaining these phenomena has been rooted in statistical physics, specifically through the lens of spin-glass theory and the Cavity Method \cite{Mezard2009, Mezard2002}.

\subsection{Phase Transitions in Statistical Physics}
According to the statistical physics perspective, the solution space of a random 3-SAT instance evolves through several distinct phases as $\alpha$ increases \cite{Mezard2002, Krzakala2007}:

\begin{enumerate}
\item \textbf{The Replica Symmetric Phase:} At low constraint densities, the problem is under-constrained. The valid solutions form a single, giant connected component. A local search algorithm can navigate between solutions by flipping a small number of variables without violating the formula.
\item \textbf{The Clustering Phase:} As the density reaches the dynamical clustering threshold ($\alpha_d \approx 3.86$ for 3-SAT), the giant connected component shatters into exponentially many disjoint ``clusters'' or states.
\item \textbf{The Rigidity Phase:} Approaching the satisfiability threshold, the solution space enters the rigidity phase. A global ``backbone'' emerges where certain variables become globally frozen, rigidly locked into specific truth values across the entire remaining valid solution space.
\item \textbf{The UNSAT Phase:} Finally, as $\alpha$ crosses the critical satisfiability threshold ($\alpha_s \approx 4.267$), the solution space vanishes entirely, and the instance becomes unsatisfiable.
\end{enumerate}

\subsection{The Need for a Discrete Microscopic Model}
While the statistical physics approach provides profound macroscopic insights, the theory relies on concepts that can be overly complex and abstract for computer science. Specifically, it lacks a clear, discrete microstructure.

In the physics literature, the shattering of the giant component at $\alpha_d$ is often described using energy landscapes, where solutions reside in ``valleys'' separated by high-energy ``ridges.'' However, in the strict Boolean logic of a SAT solver, there is no energy gradient; an assignment is either a valid solution or a violation. This thermodynamic analogy leaves an explanatory gap: what exactly are these valleys and ridges in a discrete solution space, and what is the exact mechanism that causes one giant component to shatter into exponentially many disjoint clusters?

\subsection{A Solution-Based Geometric Approach}
To bridge this gap, we propose a concrete microscopic model based directly on the topological relationships between discrete valid solutions rather than abstract energy states. By representing the entire search space as an $N$-dimensional Boolean hypercube, we can explicitly map the effects of increasing constraint density on the geometry of the solution space.

Our model establishes the following discrete topological mechanics:
\begin{enumerate}
\item \textbf{Relationship Between Solutions and Clauses:} The unconstrained search space contains $2^N$ candidate solutions. Every uniquely generated clause acts as a strict geometric filter that removes a specific subset of vertices from this hypercube.
\item \textbf{Constraint Density and Solution Count:} Increasing $\alpha$ strictly monotonically decreases the total number of valid solutions by carving void spaces into the hypercube.
\item \textbf{Cluster Formation:} Two solutions belong to the same cluster if they are connected by a Hamming distance of one. When added clauses remove all neighboring solutions that bridge two regions of the hypercube, the void space acts as an impassable barrier, isolating the regions and forming disjoint clusters.
\item \textbf{Frozen Variables:} A variable becomes locally frozen when all valid solutions representing its negation are removed from a specific cluster. It becomes globally frozen when its negation is removed from the entire hypercube.
\item \textbf{Unsatisfiability:} The UNSAT phase is reached simply when the accumulation of clauses successfully removes every single vertex from the $2^N$ hypercube.
\end{enumerate}

By defining the solution space through these strict combinatorial mechanics, this paper aims to demystify the complex phase transitions of random 3-SAT instances and provide a clear geometric explanation for the well-known ``easy-hard-easy'' computational complexity phenomenon.

\section{Definitions and Microscopic Model}

\label{sec:definition}

To construct a discrete geometric model of the solution space for random 3-SAT instances, we must first formalize the environment in which the solutions reside. Let $V = \{x_1, x_2, \dots, x_N\}$ be a set of $N$ Boolean variables. The unconstrained search space is represented as an $N$-dimensional Boolean hypercube, denoted as $\mathcal{H}^N = \{0, 1\}^N$, containing $2^N$ possible truth assignments. Let $F$ be a 3-CNF formula composed of $M$ clauses. We define $S \subseteq \mathcal{H}^N$ as the set of all valid satisfying assignments (solutions) for $F$. 

As clauses are added to $F$, specific vertices (assignments) are removed from $\mathcal{H}^N$. The resulting topology of the remaining vertices in $S$ dictates the computational hardness of the instance. We formalize this topology through the following definitions.

\subsection{Definitions}

\textbf{Definition 1 (Constraint Density).} 
The constraint density, denoted by $\alpha$, is the fundamental control parameter that dictates the structural phase transitions of the solution space. It is defined as the ratio of the number of randomly generated clauses $M$ to the number of Boolean variables $N$:
$$ \alpha = \frac{M}{N} $$
As $\alpha$ increases, the number of valid solutions $|S|$ monotonically decreases, gradually carving void spaces into the hypercube $\mathcal{H}^N$.
\vspace{1em}

\noindent \textbf{Definition 2 (Hamming Distance).} 
For any two truth assignments $s_a, s_b \in \mathcal{H}^N$, the Hamming distance $d_H(s_a, s_b)$ is the number of variables for which $s_a$ and $s_b$ possess differing truth values. Geometrically, two assignments are connected by a single edge on the $N$-dimensional hypercube if and only if:
$$ d_H(s_a, s_b) = 1 $$

\vspace{1em}
\noindent \textbf{Definition 3 (Solution Cluster).} 
A solution cluster $\mathcal{C}$ is a maximal connected component within the valid solution space $S$. Formally, $\mathcal{C} \subseteq S$ such that for any two solutions $s_i, s_j \in \mathcal{C}$, there exists a sequence of valid solutions $(s_i, s_{k_1}, s_{k_2}, \dots, s_j)$ all belonging to $\mathcal{C}$, where the Hamming distance between any two adjacent solutions in the sequence is exactly $1$. Two clusters $\mathcal{C}_1$ and $\mathcal{C}_2$ are strictly disjoint if no solution in $\mathcal{C}_1$ has a Hamming distance of $1$ to any solution in $\mathcal{C}_2$.

\vspace{1em}
\noindent \textbf{Definition 4 (Frozen Variable).} 
A variable $x_i \in V$ is considered \textbf{locally frozen} within a specific cluster $\mathcal{C}$ if it evaluates to the same truth value (either strictly $0$ or strictly $1$) for every solution $s \in \mathcal{C}$. Geometrically, this occurs when the addition of clauses has removed all neighboring solutions in $\mathcal{C}$ that represent the negation of $x_i$. 

Furthermore, a variable $x_i$ is considered \textbf{globally frozen} if it takes the exact same truth value across the entire valid solution space $S$.

\subsection{The Three-Variable Microscopic Model}

To illustrate the discrete topological mechanics defined above, we first examine the minimal complete geometric model: a 3-SAT instance with exactly three variables, $V = \{x_1, x_2, x_3\}$. The complete, unconstrained search space is the 3-dimensional Boolean hypercube $\mathcal{H}^3$, which contains exactly $2^3 = 8$ vertices (candidate solutions). In a strict 3-SAT formulation with no duplicated variables within a clause, there are exactly $2^3 = 8$ unique possible clauses. 

Because each unique clause contains exactly one literal for each of the three distinct variables, each clause acts as a perfect filter that invalidates exactly one specific vertex in $\mathcal{H}^3$. By incrementally adding uniquely generated clauses, we can observe the exact microscopic mechanisms of clustering, freezing, and unsatisfiability.

\begin{enumerate}
    \item \textbf{The Giant Component (Under-constrained):} 
    Consider an instance with a single clause, e.g., $(x_1 \lor x_2 \lor x_3)$, which falsifies exactly one assignment, $(0, 0, 0)$. The remaining $7$ valid solutions still form a single connected component $\mathcal{C}$. Because every remaining solution has at least one neighbor at a Hamming distance of $1$, a local search algorithm can traverse the entire valid space without leaving $S$. There are no frozen variables.

    \item \textbf{Shattering and Local Freezing:} 
    As the constraint density increases, the addition of specific clauses can shatter the giant component. Suppose the instance contains three clauses designed to remove the neighbors of the assignment $(0,0,0)$: $(\neg x_1 \lor x_2 \lor x_3)$, $(x_1 \lor \neg x_2 \lor x_3)$, and $(x_1 \lor x_2 \lor \neg x_3)$. These clauses remove the assignments $(1,0,0)$, $(0,1,0)$, and $(0,0,1)$ respectively. 
    
    This severs all Hamming distance bridges to $(0,0,0)$, shattering the solution space $S$ into two strictly disjoint clusters:
    \begin{itemize}
        \item $\mathcal{C}_1 = \{(0,0,0)\}$: An isolated point. Because no variable can change its value without exiting the cluster, $x_1$, $x_2$, and $x_3$ are all \textbf{locally frozen}.
        \item $\mathcal{C}_2$: A connected cluster of the remaining $4$ valid solutions. Within $\mathcal{C}_2$, all variables flip their values at least once, meaning there are $0$ frozen variables in this cluster.
    \end{itemize}

    \item \textbf{The Emergence of Global Freezing:} 
    Different clause combinations dictate whether freezing is local or global. Consider an instance with four clauses that explicitly remove all assignments where $x_1 = 0$. The remaining valid solutions are $(1,1,1)$, $(1,1,0)$, $(1,0,1)$, and $(1,0,0)$. 
    
    Geometrically, this removes an entire face of the hypercube. The $4$ remaining solutions form a single, fully connected 2D square (a single cluster). However, because $x_1 = 1$ in every remaining valid solution, $x_1$ becomes a \textbf{globally frozen} variable, acting as a rigid backbone for the entire search space.

    \item \textbf{The Unsatisfiability Threshold:} 
    If the instance continues to accumulate unique clauses, it will eventually contain all $8$ possible clauses. At this absolute maximum constraint density, all $8$ vertices of $\mathcal{H}^3$ are removed. The solution space becomes entirely empty ($S = \emptyset$), strictly crossing the unsatisfiability threshold.
\end{enumerate}

By observing the 3-variable model, we establish that the transition from a highly connected space to an unsatisfiable one is not a continuous thermodynamic decay, but a strict combinatorial process of vertex elimination that predictably shatters graphs and locks variable values.

\subsection{Extending to the $N$-Variable Problem}

While the 3-variable model perfectly isolates the mechanics of clustering and freezing, real-world SAT problems operate in a massively high-dimensional space where $N \gg 3$. To scale our microscopic model to the general $N$-variable case, we must analyze the topological tension between the expanding volume of the Boolean hypercube and the restrictive filtering of added clauses.

\subsubsection{The Necessity of the Control Parameter $\alpha$}
In an $N$-variable unconstrained space, the hypercube $\mathcal{H}^N$ contains $2^N$ vertices. Because every vertex has exactly $N$ neighbors at a Hamming distance of one, an increase in $N$ exponentially expands the volume of the space. This massive connectivity naturally resists the formation of disjoint clusters and frozen variables.

Conversely, each generated 3-SAT clause acts as a strict filter. A single 3-variable clause eliminates exactly $\frac{1}{8}$ of the unconstrained space; specifically, it removes $2^{N-3}$ assignments. As the number of clauses $M$ increases, this constant fractional carving aggressively combats the expanding volume. 

Because $M$ and $N$ exert opposing forces on the topology of the solution space, tracking them individually for large formulas is impractical. Instead, we rely on the constraint density parameter, $\alpha = \frac{M}{N}$, which perfectly captures this thermodynamic tension. Due to the massive combinatorial explosion of clause arrangements in general cases, analyzing the exact geometry becomes intractable. Therefore, as $\alpha$ increases, we shift from mapping exact topologies to tracking the \textbf{probability of satisfiability} and the probable emergence of structural anomalies.

\subsubsection{Microscopic Traps in Macroscopic Space}
The strict combinatorial logic established in our 3-variable model remains absolute in the $N$-variable general case. Because clauses in 3-SAT are strictly localized to three variables, global structural collapse can be triggered by purely localized microscopic traps. 

Assuming a random generation of clauses, we can define the absolute earliest emergence of structural phase transitions based entirely on 3-variable subsets:
\begin{enumerate}
    \item \textbf{Early Clustering ($\alpha = \frac{3}{N}$):} If an instance generates three specific clauses on the same three variables---e.g., $(\neg x_1 \lor x_2 \lor x_3)$, $(x_1 \lor \neg x_2 \lor x_3)$, and $(x_1 \lor x_2 \lor \neg x_3)$---it removes all Hamming distance bridges out of the $2^{N-3}$ subcube where $x_1=0, x_2=0, x_3=0$. Although the probability is infinitesimally small for large $N$, it proves that strictly disjoint clusters can theoretically emerge as early as $M=3$.
    \item \textbf{Early Global Freezing ($\alpha = \frac{4}{N}$):} If four specific clauses fall on $x_1, x_2$, and $x_3$ such that all combinations where $x_1=0$ are falsified, then $x_1$ is forced to be $1$. Because this restriction applies independently of the remaining $N-3$ variables, $x_1$ becomes strictly globally frozen across the entire remaining search space as early as $M=4$.
    \item \textbf{Early Unsatisfiability ($\alpha = \frac{8}{N}$):} If eight unique clauses are generated on the exact same three variables, they form a localized ``unsatisfiable core.'' It does not matter what values the other $N-3$ variables take; the entire $N$-variable formula becomes fundamentally unsatisfiable as early as $M=8$.
\end{enumerate}

\subsubsection{Absolute Structural Bounds for 3-SAT}
Based on these microscopic traps and the geometry of the Boolean hypercube, we can define the absolute probabilistic boundaries for any random 3-SAT instance. These limits exist completely independently of the average-case statistical thresholds.

\begin{itemize}
	\item \textbf{The Absolute SAT Bound ($M<8$):} 
In \emph{strict} random 3-SAT, each clause contains three \emph{distinct} variables (and is non-tautological), so it falsifies exactly $2^{N-3}$ assignments in the $N$-dimensional hypercube. Hence, for any formula with $M$ clauses, the total number of assignments ruled out is at most $M\cdot 2^{N-3}$. When $M\le 7$, this is strictly less than $2^N$, so at least one assignment remains that satisfies all clauses. Therefore every strict 3-SAT formula with $M\le 7$ is satisfiable, and for $\alpha \in \left[0,\frac{7}{N}\right]$ the satisfiability probability equals $1$.

	\item \textbf{The Minimal UNSAT Cores ($M = 8$):} In strict 3-SAT, creating an unavoidable contradiction requires at least $8$ clauses. This bound is tight: an unsatisfiable core of size $8$ is obtained by taking all $8$ distinct clauses over a fixed variable triple $(x_i,x_j,x_k)$, which forbids every assignment to that subcube. Because there are $\binom{N}{3} = \frac{N(N-1)(N-2)}{6}$ ways to select a 3-variable subset from $N$ variables, there are \textit{at least} $\frac{N(N-1)(N-2)}{6}$ distinct 8-clause configurations that immediately form a strict 3-variable unsatisfiable core. Furthermore, this constitutes only a lower bound; as variables interact across higher dimensions, the true combinatorial volume of distinct instances that become unsatisfiable at exactly $M=8$ is significantly larger.

    \item \textbf{The Maximum Clause Limit ($M_{max}$):} In strict 3-SAT (no duplicated variables within a clause), the total number of unique 3-variable combinations is $\binom{N}{3}$. Since each combination supports $8$ unique clauses, the absolute maximum number of possible unique clauses is:
    $$ M_{max} = 8 \binom{N}{3} = \frac{4}{3}N(N-1)(N-2) $$

    \item \textbf{Maximal Satisfiable Instances:} It is possible to construct highly dense instances that remain satisfiable. Consider a set of ``special clauses'' where all three variables are positive literals. There is exactly one such clause for every 3-variable combination, totaling $\frac{N(N-1)(N-2)}{6}$ special clauses. If an instance includes all possible clauses \textit{except} these special all-positive clauses, it will contain exactly:
    $$ M_{max} - \frac{N(N-1)(N-2)}{6} = \frac{7}{6}N(N-1)(N-2) \text{ clauses} $$
    Despite this extreme density, the instance remains satisfiable because the assignment where all variables are set to \texttt{False} perfectly satisfies every included clause. 
    
    Crucially, this logic generalizes to the entire hypercube. For any of the $2^N$ possible target solutions in the search space, we can construct exactly one such maximal instance by excluding only the specific $\frac{N(N-1)(N-2)}{6}$ clauses that directly falsify that target. Therefore, there are exactly $2^N$ distinct maximal satisfiable instances of this size. While the probability of randomly generating one of these specific instances is infinitesimally small, their existence proves a vast combinatorial presence at the upper bounds of satisfiability.

    \item \textbf{The Absolute UNSAT Bound:} Based on the maximal satisfiable instances, if we add even one more unique clause, it must be drawn from the excluded set of clauses, which will instantly falsify the single remaining solution. Therefore, any instance where $M > \frac{7}{6}N(N-1)(N-2)$ is strictly and absolutely unsatisfiable.
    
    \item \textbf{The Probabilistic Transition Range:} Between these two absolute limits, the satisfiability of the instance is topological and probabilistic. For the wide range of constraint density:
    $$ \alpha \in \left[ \frac{8}{N}, \frac{7}{6}(N-1)(N-2) \right] $$
    the probability of satisfiability is strictly greater than $0$. Within this range, macroscopic phase transitions---such as the shattering of the giant component at $\alpha_d$ and the average-case satisfiability threshold at $\alpha_s$---occur dynamically as the hypercube collapses.
\end{itemize}

\section{Discussion}

The microscopic geometric model presented in Section~\ref{sec:definition} provides a strict combinatorial framework for understanding random 3-SAT instances. In this section, we discuss how our discrete model contextualizes the established macroscopic theories of statistical physics and how it explicitly maps to the computational complexity observed in complete search algorithms.

\subsection{Comparing with Theories in Statistical Physics}

The analytical framework of statistical physics, particularly Replica Symmetry Breaking (RSB) and the Cavity Method, has provided deep insights into the average-case behavior of random SAT formulas. Our microscopic model does not contradict these macroscopic theories; rather, it provides a complementary, strictly Boolean lens through which to view the exact same phase transitions.

\subsubsection{Alignment of the General Macroscopic Picture}
Both statistical physics and our microscopic geometric model agree on the fundamental narrative: as the constraint density $\alpha$ increases, the geometry of the solution space undergoes dramatic, structural phase transitions. We can map the physical phases directly to our topological states:
\begin{enumerate}
    \item \textbf{The Giant Component Phase:} Corresponds to the Replica Symmetric (RS) phase. At low $\alpha$, the solution space is highly connected. In our model, this is the state where valid solutions retain a large fraction of their $N$ original Hamming neighbors, forming a navigable, continuous hypercube.
    \item \textbf{The Clustering Phase:} Corresponds to the dynamic 1RSB phase. Both frameworks agree that the giant component breaks apart into isolated sub-spaces.
\item \textbf{The Freezing Phase:} Corresponds to the rigidity phase. Variables become globally frozen, locking into specific truth values across all remaining valid solutions, rather than just locally within individual clusters.
    \item \textbf{The UNSAT Phase:} Both models agree that at a critical upper bound, the solution space vanishes entirely.
\end{enumerate}

\subsubsection{Differences in Core Mechanics and Explanations}
While the macroscopic stages align, our model diverges from statistical physics in its fundamental definitions and its description of the underlying mechanical processes. 

Statistical physics heavily relies on thermodynamic analogies, defining the problem space through \textit{energy landscapes}, partition functions, and probabilistic message passing. In that framework, clustering and freezing are the results of solutions settling into deep ``valleys'' separated by high-energy ``ridges.'' 

Conversely, our model relies strictly on a discrete Boolean topology.
\begin{itemize}
    \item \textbf{Clustering} is simply the geometric consequence of neighbor solutions being explicitly removed by clauses, severing the Hamming distance bridges between valid regions of the hypercube.    
    \item \textbf{Freezing} is the strict deterministic result of negation solutions being eliminated from a local cluster, leaving a variable with only one valid Boolean state.
\end{itemize}

\subsubsection{The Process of Change: Gradual Evolution vs. Sudden Shattering}
The most significant distinction between the two frameworks lies in how they describe the \textit{process} of these phase transitions. 

In statistical physics, the transition at the dynamic clustering threshold ($\alpha_d$) is often described as a sudden, dramatic event where the giant connected component instantly shatters into exponentially many disjoint clusters \cite{Achlioptas2021, Achlioptas2008}, with the largest cluster potentially dominating the thermodynamic state. 

From our microscopic perspective, the evolution is far more gradual and deterministic. Because each uniquely generated clause simply removes a specific $2^{N-3}$ subcube of assignments, the structural degradation happens sequentially, clause by clause. 
\begin{itemize}
    \item \textbf{Gradual Clustering:} As $\alpha$ increases, clusters do not necessarily shatter into exponentially many pieces all at once. Instead, they emerge progressively. An initial giant component might be sliced into two clusters, then three, and so on. Furthermore, because clauses are uniformly random, the slicing is topologically arbitrary; the largest cluster does not inherently dominate the algorithmic search space.
    \item \textbf{Gradual Freezing:} Similarly, the frozen fraction of variables does not undergo a sudden phase change. As more clauses are added, more negation assignments are progressively eliminated. Thus, the fraction of locally and globally frozen variables increases gradually as the hypercube is systematically carved away.
\end{itemize}

Ultimately, while statistical physics describes the aggregate thermodynamic limits of the solution space for infinite $N$, our microscopic model describes the strict, clause-by-clause geometric realities that an actual search algorithm must navigate in a finite state space.

\subsection{Comparing with the Satisfiability Threshold Conjecture}

The Satisfiability Threshold Conjecture is one of the most studied phenomena in the computational complexity of random 3-SAT. Our microscopic model provides a discrete, structural explanation for the probabilistic behaviors described by this conjecture.

\subsubsection{Alignment on the General Phase Transition}
The overarching picture presented by both our model and the conjecture is highly aligned: as the constraint density $\alpha$ increases, the random formula undergoes a fundamental phase transition from a state of satisfiability (SAT) to unsatisfiability (UNSAT). The solution space degrades until it vanishes.

\subsubsection{Divergence in Description: Average-Case vs. Absolute Bounds}
The difference between the two frameworks lies in how they describe the boundaries of this transition. The Satisfiability Threshold Conjecture relies on asymptotic limits ($N \to \infty$) to describe an average-case behavior, rigorously shown to exhibit a sharp threshold \cite{Friedgut1999, Achlioptas2021}, whereas our microscopic model establishes absolute structural boundaries that hold for any finite $N$.

\begin{itemize}
    \item \textbf{The Conjecture's Description:} The conjecture states that for random 3-SAT as $N \to \infty$, there exists a sharp, constant threshold $\alpha_s$. If the constraint density $\alpha < \alpha_s$, the formula is \textit{almost surely} satisfiable (the probability of satisfiability approaches $1$). Conversely, if $\alpha > \alpha_s$, the formula is \textit{almost surely} unsatisfiable (the probability approaches $0$).
    \item \textbf{Our Model's Description:} Instead of a single point of failure, our model identifies two absolute structural transitions that bracket the conjecture's threshold $\alpha_s$:
    \begin{enumerate}
        \item At $\alpha = \frac{8}{N}$, the system transitions from being \textit{absolutely satisfiable} to \textit{possibly satisfiable}.
        \item At $\alpha = \frac{7}{6}(N-1)(N-2)$, the system transitions from being \textit{possibly satisfiable} to \textit{absolutely unsatisfiable}.
    \end{enumerate}
    The experimental average-case threshold $\alpha_s$ strictly resides within this massive combinatorial range $\left( \frac{8}{N}, \frac{7}{6}(N-1)(N-2) \right)$.
\end{itemize}

\subsubsection{The Population Distribution and the Meaning of ``Almost Surely''}
Our model perfectly elucidates the mechanical reality behind the conjecture's use of the phrase ``almost surely.'' 

Within the bounds of our two absolute transitions, there exists a population distribution of SAT and UNSAT instances. For small finite values of $N$, this distribution is relatively flat. As $N \to \infty$, the combinatorial volume of instances near the critical density explodes, causing the probability distribution to sharpen into an extremely narrow peak exactly at $\alpha_s$. 

However, our model proves that no matter how large $N$ becomes, the absolute boundaries do not collapse into $\alpha_s$, and the tails of the distribution never truly vanish:
\begin{itemize}
    \item Even at an infinitesimally small constraint density ($\alpha = \frac{8}{N} \ll \alpha_s$), there strictly exist at least $\frac{N(N-1)(N-2)}{6}$ distinct instances that are fundamentally unsatisfiable due to microscopic cores.
    \item Even at a massively over-constrained density ($\alpha = \frac{7}{6}(N-1)(N-2) \gg \alpha_s$), there strictly exist exactly $2^N$ distinct instances that remain perfectly satisfiable.
\end{itemize}

Therefore, an ``absolutely sure'' structural rule---where the probability of satisfiability is strictly $1$ or strictly $0$---only exists outside our defined combinatorial boundaries ($\alpha < \frac{8}{N}$ and $\alpha > \frac{7}{6}(N-1)(N-2)$). Within these bounds, while the vast statistical majority of random instances obey the sharp average-case threshold $\alpha_s$, there are a strictly quantifiable number of instances that deeply violate the conjecture. This geometric reality mathematically justifies why the Satisfiability Threshold Conjecture can only claim that instances are ``almost surely'' SAT or UNSAT near the threshold, reserving absolute certainty solely for the absolute structural limits.

\subsection{Explaining the Easy-Hard-Easy Phenomenon}

The structural phase transitions defined in our geometric model directly govern the computational complexity of solving random 3-SAT instances, famously characterized as the ``easy-hard-easy'' phenomenon \cite{Mitchell1992, Cheeseman1991}. Because the concept of algorithmic ``hardness'' is strictly dependent on the search method utilized, we analyze this phenomenon through the lens of a Depth-First Search (DFS) branching algorithm. To isolate the effect of constraint density, we hold the number of variables $N$ constant.

\subsubsection{The First ``Easy'' Phase ($\alpha \ll \alpha_s$): The Abundance of Paths}
In the under-constrained regime, the instance is almost certainly satisfiable. For a DFS algorithm, the easiness of a SAT instance is inversely proportional to the number of valid solutions remaining in the hypercube. 

If an instance contains $N_{sol}$ solutions distributed across the $2^N$ possible assignments, we can represent this density as $N_{sol} = \frac{2^N}{r}$, where $r$ is a real-valued \textit{sparsity metric}. On average, a basic DFS algorithm will need to explore $r$ branches (or paths) before successfully encountering a valid assignment. 

Based on our microscopic model, every added clause randomly eliminates $\frac{1}{8}$ of the unconstrained space. Therefore, the expected number of solutions $N_{sol}$ for a given number of clauses $M$ is:
$$ N_{sol} = 2^N \left( \frac{7}{8} \right)^M $$
Equating our two expressions for $N_{sol}$, we can directly solve for the expected search effort $r$:
$$ r = \left( \frac{8}{7} \right)^M $$
Substituting $M = \alpha N$ and taking the natural logarithm, we establish the strict mathematical relationship between the search effort and the constraint density $\alpha$:
$$ \ln(r) = N \alpha \ln\left(\frac{8}{7}\right) $$
This equation clearly demonstrates that for a fixed number of variables $N$, the logarithmic expected search effort $\ln(r)$ has a strict linear relationship with the constraint density $\alpha$. 

When $\alpha$ is small, $r$ is extremely small. Because the target density is so high, the DFS algorithm easily drops into the giant connected component and finds a solution with minimal backtracking, characterizing the first ``easy'' phase.

\subsubsection{The Second ``Easy'' Phase ($\alpha \gg \alpha_s$): Removed Variables and Shallow Search}
In the massively over-constrained regime, the instance is almost certainly unsatisfiable. Here, the algorithm does not search for a solution, but rather searches to prove that no solution exists. The efficiency of this proof relies on a geometric concept we define as \textbf{removed variables}.

As established in Section~\ref{sec:definition}, a variable becomes locally or globally \textit{frozen} when all solutions representing its negation are removed from the space. As $\alpha$ increases further, the continuous accumulation of clauses eventually removes the solutions for \textit{both} the positive and negative literals of a given variable within a subcube. When this occurs, the variable itself is effectively \textit{removed} from the valid state space. No matter whether the DFS assigns \texttt{True} or \texttt{False} to this variable, it immediately violates a clause. 

Similar to the average frozen fraction of variables, the average fraction of removed variables increases monotonically with $\alpha$. 
\begin{itemize}
    \item \textbf{High Depth Search (Low $\alpha$):} If the number of removed variables is small, a DFS algorithm must descend deep into the decision tree---assigning values to many variables---before encountering a contradiction that forces it to stop and backtrack.
    
    \item \textbf{Shallow Depth Search (High $\alpha$):} As $\alpha$ grows large, the number of removed variables dominates the space. If $x_1$ is a removed variable, the DFS will immediately discover that both $x_1 = 1$ and $x_1 = 0$ lead to unavoidable conflicts. The search completely halts at a depth of $1$, instantly proving unsatisfiability.
    
\end{itemize}
Therefore, highly constrained UNSAT instances often become “easy” in practice: the high density of removed variables tend to induce early contradictions, which prunes the search tree and can drastically reduce the depth explored before a refutation is found.

This geometric topology also explains the experimental phenomenon observed by Selman et al \cite{Selman1996}, wherein the computational hardness of the strictly satisfiable instances is shown to flatten and even decrease as $\alpha$ increases past the critical threshold. While the abundance of paths explains the easiness at low constraint densities ($\alpha \ll \alpha_s$), the rare satisfiable instances that survive at high constraint densities ($\alpha > \alpha_s$) become easier for a completely different geometric reason: the space is dominated by globally and locally frozen variables. If a DFS algorithm branches incorrectly by assigning the negated value to a frozen variable, the hypercube geometry forces an unavoidable contradiction at a very shallow depth.  This high density of frozen variables acts as a rigid funnel, aggressively pruning the search tree and rapidly forcing the algorithm down the narrow, remaining valid paths without requiring deep, exhaustive backtracking.

\subsubsection{The ``Hard'' Peak ($\alpha \approx \alpha_s$): The Composite Ensemble Effect}
The peak computational hardness observed in experimental data does not represent a single, uniform geometric state, but rather the composite average of a mixed ensemble. Near the critical threshold $\alpha_s$, approximately half of the randomly generated instances are satisfiable, and half are unsatisfiable. The ``easy-hard-easy'' curve is the weighted average of these two distinct structural populations as their respective probabilities shift.

As the constraint density $\alpha$ increases across the transition range, the algorithmic hardness is driven by two competing geometric effects:

\begin{enumerate}
	\item \textbf{Satisfiable Instances (Increasing Hardness, Decreasing Frequency):} As $\alpha$ grows, $\ln(N_{sol})$ shrinks toward $0$. The solutions become exceptionally rare and are often isolated within disjoint clusters protected by locally frozen variables. This forces the DFS algorithm into deep, exhaustive backtracking, increasing the computational cost. However, the statistical percentage of instances that are satisfiable drops precipitously toward $0$.
	\item \textbf{Unsatisfiable Instances (Decreasing Hardness, Increasing Frequency):} Conversely, as $\alpha$ grows, the fraction of \textit{removed variables} increases. This allows the DFS algorithm to detect unavoidable conflicts at progressively shallower depths, decreasing the computational cost of proving unsatisfiability. Concurrently, the statistical percentage of instances that are unsatisfiable rises toward $100\%$.
\end{enumerate}

The macroscopic ``hard peak'' emerges as the strict mathematical intersection of these trends. Initially, the rising search cost of the SAT instances dominates the average, pulling the overall computational hardness up. As $\alpha$ crosses the threshold, the rapidly increasing frequency of the progressively easier UNSAT instances begins to dominate the ensemble, pulling the overall average computational hardness back down.

Because the raw computational cost of finding a deeply hidden solution in a SAT instance may not perfectly equal the cost of proving a shallow contradiction in an UNSAT instance at $\alpha_s$, the absolute peak of the experimental hardness curve is located near the critical threshold, but naturally skews toward the side of the distribution with the higher absolute computational cost.

\section{Conclusion}
The structural phase transitions and computational complexity of random 3-SAT instances have traditionally been analyzed through the macroscopic, continuous frameworks of statistical physics. In this paper, we proposed a discrete, microscopic geometric model that strictly grounds these phenomena in the combinatorial topology of an $N$-dimensional Boolean hypercube. By defining the problem space purely in terms of valid solutions rather than abstract energy states, we established a clear, deterministic mechanism for how the accumulation of constraints systematically degrades the solution space.

Our geometric framework demonstrates that the dramatic phase transitions observed in random 3-SAT---such as clustering and freezing---are not the result of sudden thermodynamic shattering, but rather the gradual, clause-by-clause elimination of specific vertices and Hamming distance bridges. Through this microscopic lens, we derived absolute structural boundaries for the problem. We identified \textit{at least} $\frac{N(N-1)(N-2)}{6}$ minimal unsatisfiable cores at the low constraint density limit of $\alpha = \frac{8}{N}$, and $2^N$ maximal satisfiable instances at the extreme density limit of $\alpha = \frac{7}{6}(N-1)(N-2)$. The existence of these massive combinatorial populations mathematically demystifies the Satisfiability Threshold Conjecture, proving why the average-case transition at $\alpha_s$ represents an ``almost sure'' statistical dominance rather than a strict structural absolute.

Furthermore, this topological mapping provides a direct, mechanical explanation for the well-known ``easy-hard-easy'' algorithmic complexity curve, revealing it as a composite ensemble effect governed by the underlying hypercube geometry: \begin{enumerate} 
	\item \textbf{The under-constrained ``easy'' phase ($\alpha \ll \alpha_s$)} is driven by an exponential abundance of valid paths, allowing solvers to quickly locate solutions with minimal branching.
	\item \textbf{The threshold ``hard'' peak ($\alpha \approx \alpha_s$)} occurs exactly when the space lacks both the abundance of paths to finish early and the structural rigidity to fail fast. Solvers are trapped in deep, disjoint clusters dominated by locally frozen variables, necessitating exhaustive backtracking. Macroscopically, this peak represents the weighted ensemble average of increasingly hard, vanishingly rare satisfiable instances and progressively easier unsatisfiable instances. 
	\item \textbf{The over-constrained ``easy'' phase ($\alpha \gg \alpha_s$)} is driven by two distinct mechanisms. For the vast statistical majority of instances (which are unsatisfiable), it is driven by a high density of \textit{removed variables}---variables for which both positive and negative assignments are entirely eliminated---forcing immediate contradictions. For the exceptionally rare instances that remain satisfiable, a massive density of \textit{frozen variables} aggressively funnels the search down narrow valid paths. Both mechanisms prohibit deep branching and keep the search tree extremely shallow.
\end{enumerate}

Ultimately, by translating the theories of statistical physics into the discrete language of vertices, clusters, frozen variables, and removed variables, this microscopic model bridges the gap between theoretical phase transitions and the tangible mechanics of complete search algorithms. Understanding these strict topological bounds not only clarifies the fundamental nature of the satisfiability threshold but also provides a rigorous geometric foundation for engineering more advanced branching and conflict-driven heuristics in modern SAT solvers.

\section*{AI Use Disclosure}
This manuscript benefited from the use of generative AI tools to assist with language-level editing, including improving clarity, grammar, and overall presentation. The AI tools were not used to generate new scientific claims, derive results, or interpret data. All ideas, analyses, conclusions, and original contributions are solely the work of the author, who reviewed and approved the final text.


\begin{thebibliography}{99}

\bibitem{Biere2021}
Biere, A., Heule, M. J. H., van Maaren, H., \& Walsh, T. (Eds.). (2021). 
\textit{Handbook of Satisfiability} (2nd ed., Vol. 336, Frontiers in Artificial Intelligence and Applications). 
IOS Press.

\bibitem{Mezard2009}
M\'{e}zard, M., \& Montanari, A. (2009). 
\textit{Information, Physics, and Computation}. 
Oxford University Press.

\bibitem{Monasson1999}
Monasson, R., Zecchina, R., Kirkpatrick, S., Selman, B., \& Troyansky, L. (1999). 
\textit{Determining computational complexity from characteristic `phase transitions'.} 
Nature, 400(6740), 133-137.

\bibitem{Mezard2002}
M\'{e}zard, M., Parisi, G., \& Zecchina, R. (2002). 
\textit{Analytic and algorithmic solution of random satisfiability problems.} 
Science, 297(5582), 812-815.

\bibitem{Krzakala2007}
Krz\k{a}ka\l{}a, F., Montanari, A., Ricci-Tersenghi, F., Semerjian, G., \& Zdeborov\'{a}, L. (2007).
\textit{Gibbs states and the set of solutions of random constraint satisfaction problems}.
Proceedings of the National Academy of Sciences, 104(25), 10318-10323.

\bibitem{Achlioptas2021}
Achlioptas, D. (2021). 
\textit{Random satisfiability.} 
In \textit{Handbook of Satisfiability} (pp. 437-462). IOS Press.

\bibitem{Achlioptas2008}
Achlioptas, D., \& Coja-Oghlan, A. (2008).
\textit{Algorithmic barriers from phase transitions}.
In \textit{2008 49th Annual IEEE Symposium on Foundations of Computer Science} (pp. 793-802). IEEE.

\bibitem{Friedgut1999}
Friedgut, E. (1999). 
\textit{Sharp thresholds of graph properties, and the k-sat problem.} 
Journal of the American Mathematical Society, 12(4), 1017-1054.

\bibitem{Mitchell1992}
Mitchell, D., Selman, B., \& Levesque, H. (1992). 
\textit{Hard and easy distributions of SAT problems.} 
In Proceedings of the 10th National Conference on Artificial Intelligence (AAAI), 459-465.

\bibitem{Cheeseman1991}
Cheeseman, P., Kanefsky, B., \& Taylor, W. M. (1991). 
\textit{Where the really hard problems are.} 
In Proceedings of the International Joint Conference on Artificial Intelligence (IJCAI), 331-337.

\bibitem{Selman1996}
Selman, B., Mitchell, D. G., \& Levesque, H. J. (1996).
\textit{Generating hard satisfiability problems}.
Artificial Intelligence, 81(1-2), 17-29.

\end{thebibliography}
\end{document}